\documentclass{article}
\usepackage{mrs2005,epsfig}
\def\simlt{\lower.5ex\hbox{$\; \buildrel < \over \sim \;$}}
\def\simgt{\lower.5ex\hbox{$\; \buildrel > \over \sim \;$}}

\begin{document} 
\title{GOODS, UDF, and the evolution of early-type galaxies}
  
\author{I. Ferreras$^{1,2}$, T. Lisker$^{1,3}$, A. Pasquali$^1$, C.~M. Carollo$^1$, S.~J. Lilly$^1$
        and B. Mobasher$^4$}
\affil{1. Institut f\"ur Astronomie, ETH-Z\"urich, CH-8093 Z\"urich, Switzerland\\
       2. Dept. Physics \& Astronomy, UCL. Gower Street, London WC1E 6BT, United Kingdom\\
       3. Astronomisches Institut, Universit\"at Basel, Venusstrasse 7, CH-4102 Binningen, Switzerland\\
       4. STScI, 3700 San Martin Drive, Baltimore, MD 21218, USA\\
}

\begin{abstract} 
The HST/ACS images of GOODS-South have been used to select a sample of
early-type galaxies, based on morphology and on the Kormendy relation.
The classification scheme does not use galaxy colours, hence it does not bias
against young stellar populations. The 249 galaxies ($i_{\rm AB}<24$)
paint a complex formation picture. Their stellar populations show
gradients which readily rule out a large range of ages among and
within galaxies. On the other hand, there is a decrease in
the comoving number density, which suggests a strong bias
when comparing local and distant early-type galaxies. This bias can be caused
either by a significant fraction of non early-type progenitors or by a 
selection effect (e.g. dust-enshrouded progenitors). The deep images
of the UDF were used to determine the structural properties of some of
these galaxies. Regarding the distribution of disky/boxy isophotes, we
do not find large differences with respect to local systems. However,
some early-types with the standard features of a red and dead galaxy
reveal interesting residuals, possible signatures of past merging
events.
\end{abstract} 
 
\section{Introduction} 
The star formation history of a galaxy is strongly dependent on its
morphology, a mechanism which is well-know since the early days of
observational cosmology. An accurate classification of morphologies is
therefore key to a complete picture of galaxy formation and
evolution. Of all morphologies, early-type systems dominate the
high-mass end and pose one of the most intriguing riddles to our
understanding of structure formation. Their old stellar components and
tight scaling relations are in remarkable contrast with the extended
assembly histories expected for such objects within the current
paradigm of galaxy formation (e.g. White \& Rees 1978).  The inherent
hierarchical nature of structure formation implies the more massive
systems assemble at later times.  The presence of massive distant red
galaxies at high redshift (F\"orster-Schreiber, et al. 2004) or the
peculiar abundance ratios in the stellar populations of early-type
galaxies (Thomas, 1999) can only be reconciled if the
``baryon physics'' of gas collapse, star formation and feedback is not
accounted for in an accurate way, so that the assembly and the
formation histories are decoupled. Indeed, the ``downsizing''
effect observed in the formation of spheroidal galaxies (e.g. Treu, et
al. 2005) implies that the baryon physics introduces an ``inverted
hierarchy''.  The determination of the star formation and assembly
histories of early-type galaxies thereby constitute one of the
cornerstones of observational cosmology.

This contribution presents work done on the selection of early-type
galaxies in one of the deepest and widest surveys currently available:
the HST/ACS images of the GOODS survey (Giavalisco et al. 2004).  The
depth and spatial resolution of the images enabled us to perform a
careful morphological classification.  The deeper $12$~arcmin$^2$
Ultra Deep Field (Beckwith, et al. 2005, in preparation) overlaps with
the GOODS/CDFS region, and the spheroidal galaxies in this field can
be explored in more detail. We present here some highlights
of the sample.  We refer interested readers to Ferreras et al. (2005)
for the details of the analysis of the GOODS/CDFS sample and to Pasquali et
al. (2005) for details of the UDF sample. A concordance $\Lambda$CDM
cosmology ($\Omega_\Lambda =0.7$, $\Omega_{\rm m}=0.3$, 
$H_0=70$~km s$^{-1}$ Mpc$^{-1}$) is used throughout this paper.

\section{Classification scheme}
In order to collect a sample of early-type galaxies, we started with a pre-selection
of the candidates. Figure~1A shows the values of concentration (defined as in
Bershady et al. 2000) and $M_{20}$ (Lotz et al. 2004)
for all source detections with $i_{\rm AB}\leq 24$. The galaxies are distributed
in this $C-M_{20}$ space along a ridge which shows two slopes, with a knee at a 
concentration $C\sim 2.4$. Simulations show that this is the trend expected
for a range of Sersic surface brightness profiles with the most concentrated
indices (n$\sim 3-4$) living in the upper-right region of the diagram.
Early-type galaxies have higher concentrations, and the varying slope suggests
a transition between the steep R$^{1/4}$ profiles of elliptical galaxies and
the exponential profiles of disks. Objects with concentration $C>2.4$ were
pre-selected and visually inspected by four of us (IF, TL, CMC and SJL). 
The final sample of visually classified early-types comprise 380 objects.
However, a visually classified sample will be contaminated by late-type 
galaxies which appear concentrated. Such is the case for bright knots of
star formation within a galaxy with an overall faint surface brightness.

Our main goal is to select a sample that resemble today's elliptical
galaxies. Needless to say, such a sample is affected by the 
progenitor bias (van~Dokkum \& Franx 2001) which implies
that stars in galaxies with a non early-type morphology can end up in
elliptical galaxies by $z=0$. Other catalogues of distant early-type galaxies
suffer from other types of biases. For instance, the large sample of COMBO~17 galaxies
(Bell et al. 2004) specifically target red galaxies. It is 
a combination of these different selection processes that allow us to
obtain a complete picture of the evolution of spheroidal galaxies.

For our purposes, the Kormendy relation (KR; 1977) acts as a filter.
It is one of the projections of the Fundamental Plane and does not
require spectroscopy. The KR has been found to hold in early-type
galaxies out to moderate redshifts (La~Barbera et al. 2003).  The KR
of the selected candidates is shown in figure~1B\footnote{The figure
shows the B-band surface brightness in the rest frame, including the
$(1+z)^4$ cosmological dimming term.}.  Filled and open circles
correspond to galaxies with red/blue colours,
respectively\footnote{The separation between red and blue galaxies is
based on the best fit from the photometric redshift
catalogue. Roughly, red galaxies have colours characteristic of local
E/S0/Sa galaxies. See Ferreras et al. (2005) for details.}.  The
selected galaxies are shown with respect to their rest frame B-band
surface brightness. Redshifts for the K-corrections were obtained from
the photometric redshift catalogue of Mobasher et
al. (2004)\footnote{One third of the galaxies have spectroscopic
redshifts from the ESO/VLT follow-up observations (Le~F\`evre et
al. 2004; Vanzella et al. 2005).}. These galaxies are at redshifts
$z\simlt 1$, corresponding to a lookback time $\Delta t\simlt 8$~Gyr.
In order to compare them with the local KR -- shown as a solid line,
including the scatter as dashed lines -- we must consider that the
stellar populations will fade as the galaxies evolve with time.

Hence, those sources falling below the local relation (solid
line) will ``drift away'' from the Kormendy relation and can thus be
rejected as galaxies that will not evolve {\sl by stellar evolution}
into this relation. The final sample comprises 249 galaxies over the
160~arcmin$^2$ field of view. Figure~2 shows the sample properties:
The $V-i$ colour, half-light radius and rest-frame $V$-band absolute
luminosity are shown as a function of redshift. The lines in the 
left panel are predictions from the population synthesis model
GALAXEV (Bruzual \& Charlot 2003) assuming an
exponentially decaying star formation rate at fixed (solar) metallicity,
with timescales 0.5 (solid), 1 (dashed) and 8~Gyr (dotted). The filled and
hollow circles represent red/blue galaxies according to integrated
photometry. Most of the early-type galaxies (selected irrespective of
colour) are old, passively evolving systems, with just 20\% of
galaxies with blue colours which could represent either a later
formation epoch or a secondary episode of star formation. The bluer
galaxies tend to be smaller and fainter. It is interesting to notice
that most of the candidates rejected by the ``Kormendy filter'' described above
are small objects at redshifts $z<0.5$. Splitting the sample about the
median redshift $z=0.7$, one finds that the fraction of blue early-types
decreases from 35\% at $z>0.7$ to 10\% at $z<0.7$.

\begin{figure}  
\begin{center}
{\leavevmode
\epsfxsize=.48\textwidth \epsfbox{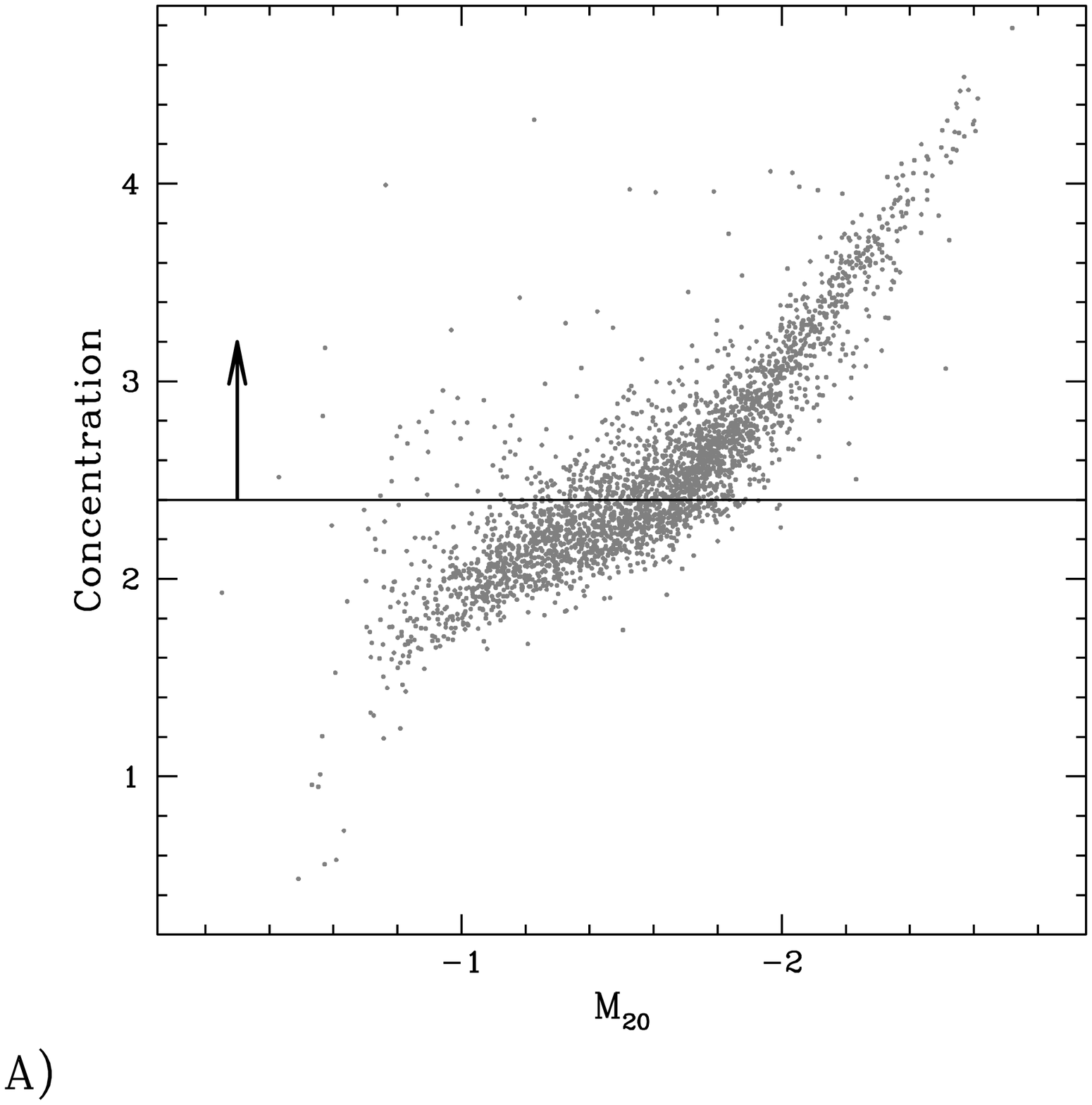} \hfil
\epsfxsize=.48\textwidth \epsfbox{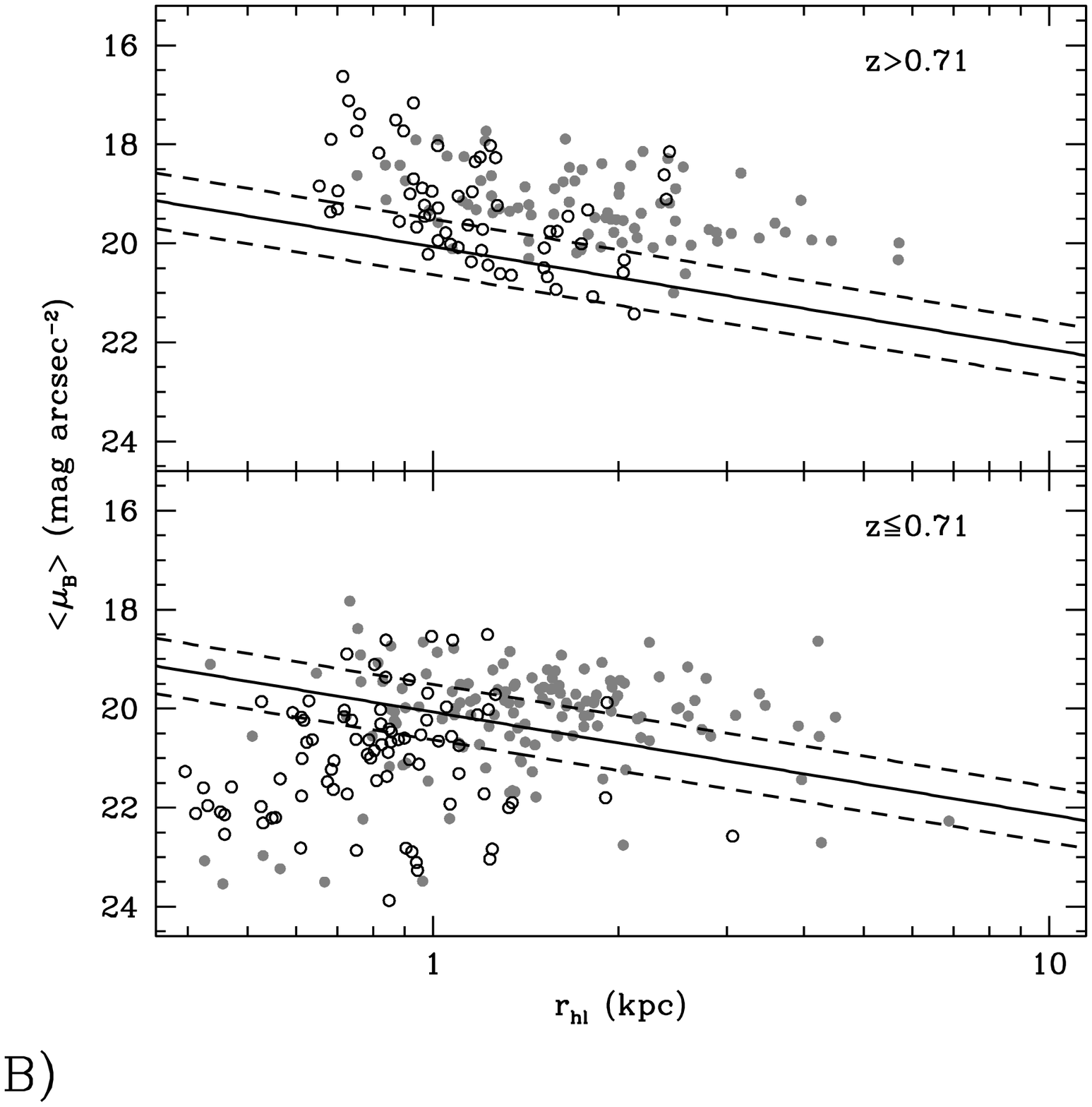}
}
\end{center}
\caption{{\bf Selection of early-type galaxies.} {\sl LEFT:} Sample pre-selection
based on the surface brightness distribution. Galaxies with $C>2.4$ were visually
inspected. {\sl RIGHT:} The Kormendy relation is shown for the galaxies visually
classified as early-types. The local Kormendy relation is shown as a solid line,
including the scatter (dashed lines).}
\label{fig:selection}
\end{figure}

\section{The formation of early-type galaxies}

Despite the small field of view (160~arcmin$^2$) one can set
constraints on the evolution of the number density of early-type
galaxies out to $z\sim 1$. Figure~3A shows the comoving number density
as a function of redshift, with the error bars including both Poisson
noise and the effect of cosmic variance. The latter is estimated from
Somerville et al.  (2004) for the volumes probed by the sample,
assuming a correlation function $\xi (r)=(r_0/r)^\gamma$, with
$r_0=5h^{-1}$~Mpc and $\gamma =1.8$. The dots comprise all galaxies
with absolute magnitude $M_{B,\star}(z)\pm 1$~mag, where
$M_{B,\star}(z)$ gives the rest-frame B-band absolute luminosity of a
typical galaxy at a given redshift. It is computed by constraining
$M_{B,\star}(z=0)$ to the local luminosity function and using a
0.5~Gyr exponentially decaying SFR along with the Bruzual \& Charlot
(2003) models to track the evolution with redshift The shaded area
illustrates the point that a zero-evolution scenario is still
compatible with the data. However, if we include the $z=0$ point
corresponding to the local density of E/S0s from Marzke et al. (1998), 
we find a significant evolution between $z=0$ and
$1$. Three lines sketch a simple power law behaviour of the density:
$\phi_\star(z)\propto (1+z)^{-\beta}$, with $\beta$ in the range 2-4.

\begin{figure}  
\begin{center}
\epsfig{figure=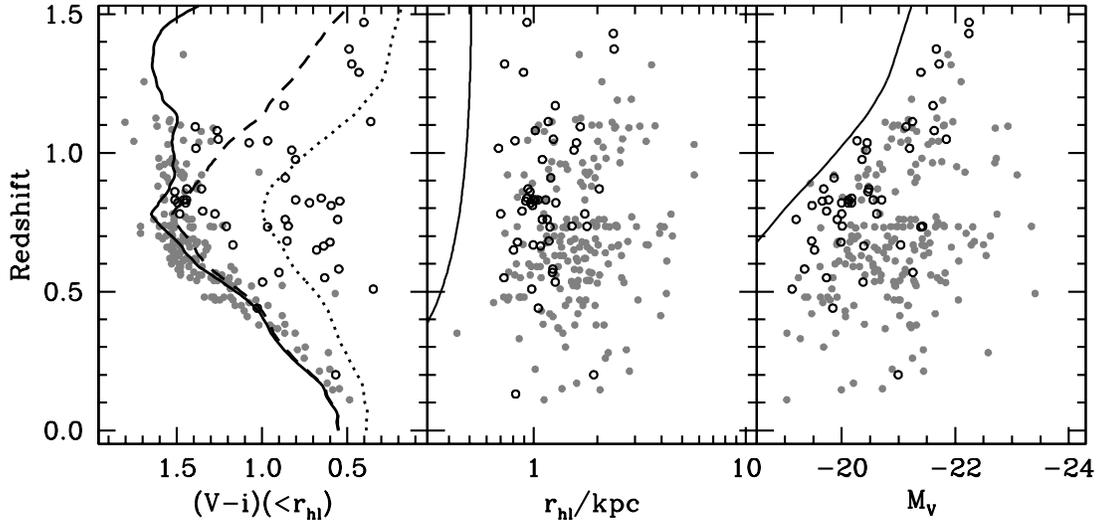,height=8.5cm}
\end{center}
\caption{{\bf GOODS-S Early-type sample.} The colour, size and absolute luminosity are
shown as a function of redshift. The solid lines in the middle and right panels
are limiting values from the detection process. The lines in 
the left panel represent an exponentially decaying star formation rate (solar
metallicity), started at $z_F=5$ with a decay timescale of 0.5 (solid), 1 (dashed) and
8~Gyr (dotted). Bruzual \& Charlot (2003) models are used.}
\label{fig:sample}
\end{figure}

The depth and spatial resolution of the ACS images allow us to perform a 
resolved colour analysis of the sample. We find a remarkable correlation between
the integrated colour and the colour gradient, so that red galaxies tend to 
have red cores and vice-versa. The large range spanned
in lookback times in our samples can be used to explore the mechanism behind the
colour gradients. Selecting the subsample of galaxies with red cores, we show 
in figure~3B the redshift evolution of the $V-i$ colour gradient (top) and scatter
(bottom). The lines follow the evolution of simple models
calibrated to the local ($B-R$) colour gradients from the sample of 
Peletier et al. (1990).
The solid line assumes the colour gradients are explained by a range of metallicities,
whereas the dashed lines assumes a range of ages. As expected, the age-sequence
evolves very quickly: In this model, the blue, outer regions have younger stars
which evolve faster than the inner parts, increasing the slope with redshift.
The metallicity-model assumes that the blue, outer regions have similar ages
but lower metallicities. This implies the inner and outer regions evolve
''in synch'', resulting in a milder evolution of the colour gradient.
The data are compatible with a pure metallicity sequence. The scatter
does not evolve with redshift. Figure~3A and 3B
show a very interesting dichotomy. The photometric properties (3B) are typical of systems
formed very early, followed by passive evolution, whereas the number density (3A) suggests
a significant evolution. 

\begin{figure}  
\begin{center}
{\leavevmode
\epsfxsize=.48\textwidth \epsfbox{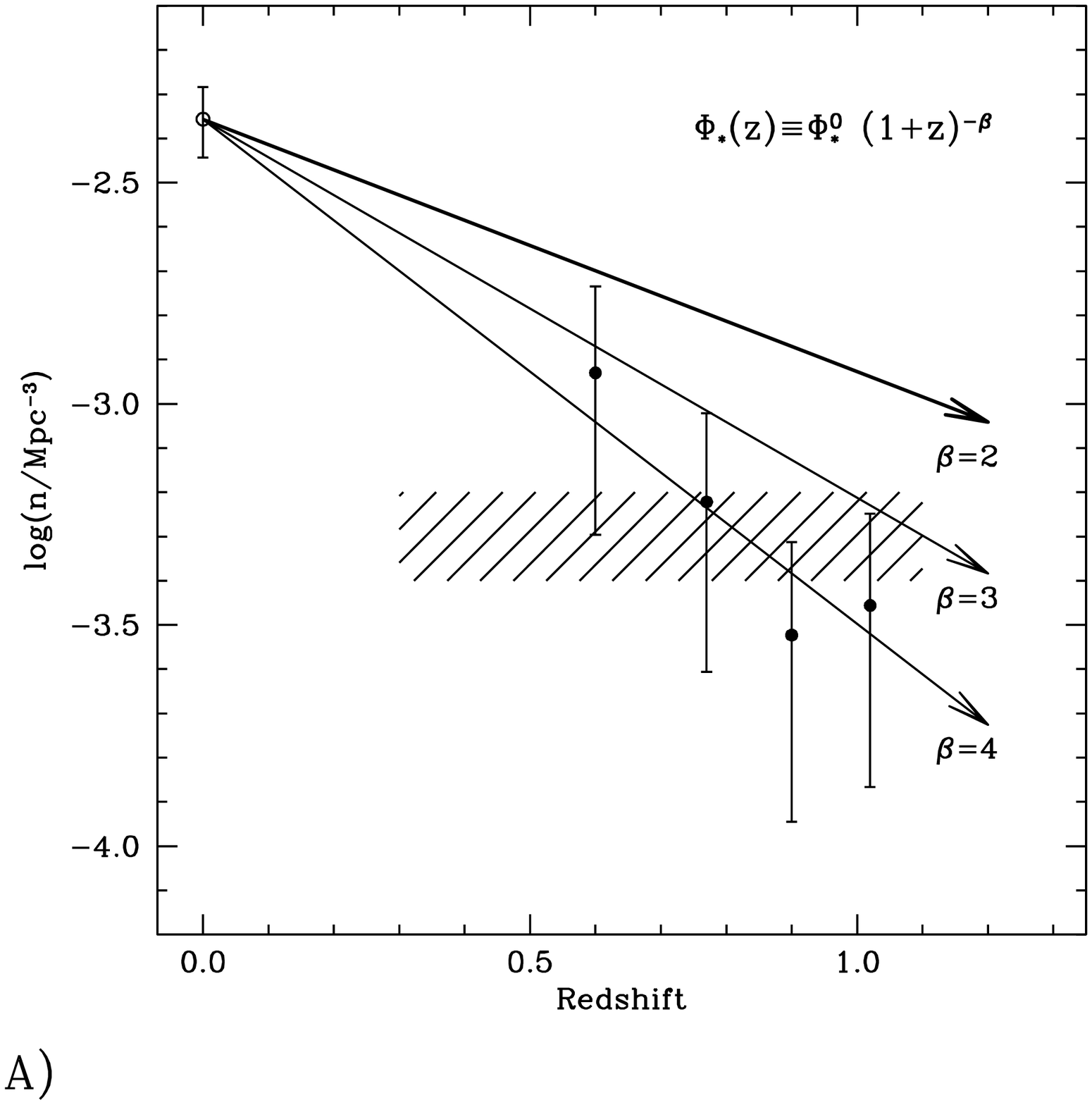} \hfil
\epsfysize=.48\textwidth \epsfbox{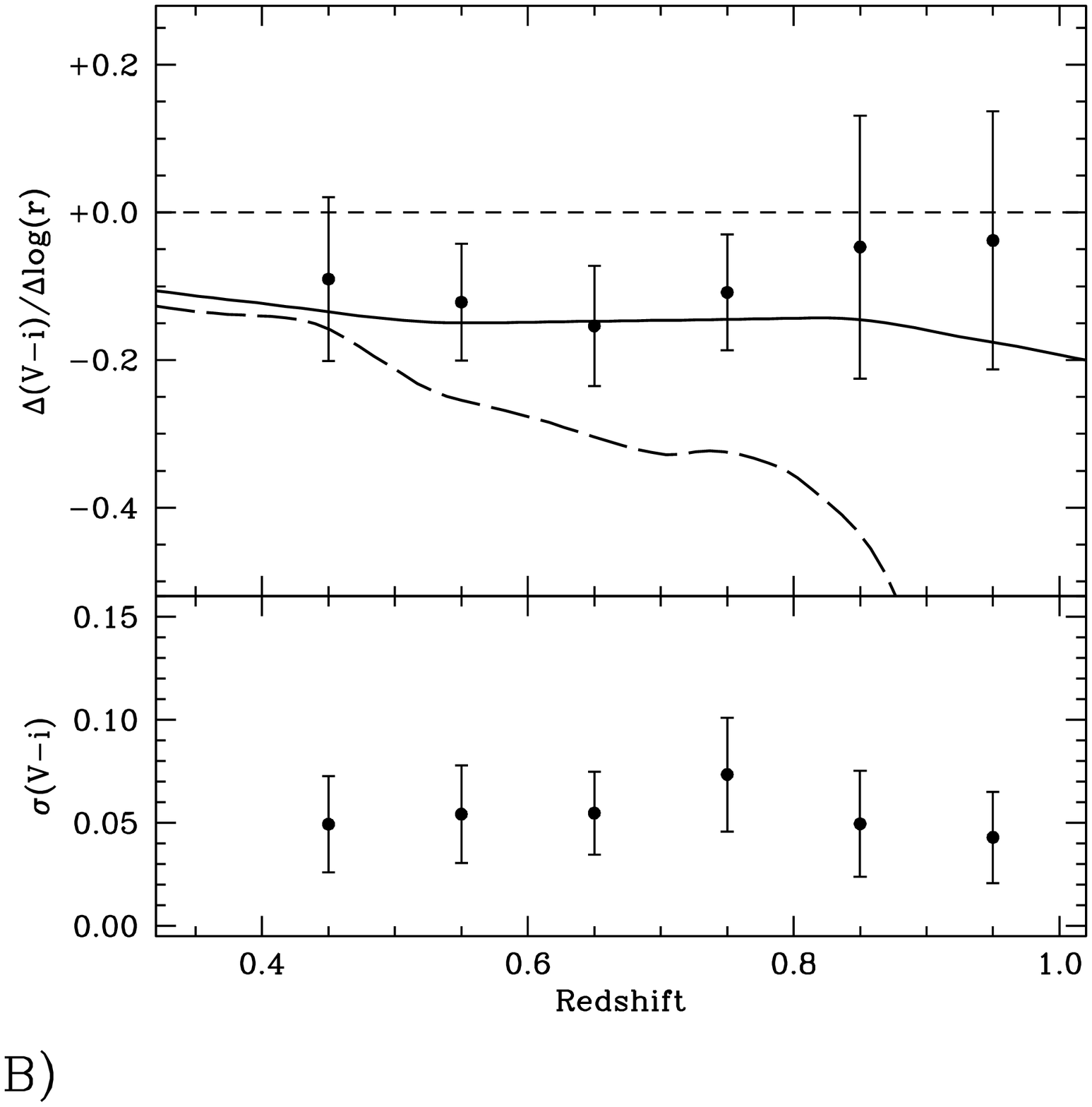}
}
\end{center}
\caption{{\bf The evolution of early-type galaxies.} {\sl LEFT:} Evolution of the
comoving number density compared to simple power laws. The $z=0$ point corresponds
to the local E/S0 luminosity function (Marzke et al. 1998). The error bars take into
account Poisson noise and cosmic variance. {\sl RIGHT:} Evolution of the colour gradient
and scatter. The solid (dashed) line corresponds to a model that relates the {\sl local}
colour gradient of ellipticals to a range of metallicities (ages).}
\label{fig:results}
\end{figure} 

\section{UDF and the structure of distant early-type galaxies}
The depth of the UDF images allowed us to explore in detail the
surface brightness distribution of some of the early-type galaxies
from the GOODS sample. Figure~4A shows the structural
parameter $\langle a_4/a\rangle$ which measures the deviation of the isophotes from
pure elliptical shapes. Positive/negative values of $\langle a_4/a\rangle$ imply
disky/boxy isophotes, respectively. Our UDF sample is
shown as black dots with respect to half-light radius, ellipticity or
rest-frame B-band absolute luminosity. The grey dots correspond to 
the local sample of Bender et al. (1989). The figure suggests an early and
uneventful morphological evolution of early-type galaxies. Similarly
to nearby elliptical galaxies, our moderate redshift sample also 
correlates diskyness/boxyness with faint/bright luminosities, respectively.

However, a detailed analysis of the isophotes reveals some interesting
features: figure~4B shows two of the 18 spheroidal galaxies
observed. Their $V-i$ colour maps are shown in the left panels. Both
objects are typical ``red and dead'' galaxies with red cores. The
central panels give the best fit for a homogeneous surface brightness
distribution ($i$ band), using the IRAF task ELLIPSE (Jedrzejewski
1987). The right panels are the residuals with respect to these fits.
While J033244.09-274541.5 (top) shows no significant residual,
UDF~2387 presents a remarkable signature suggesting a recent dynamical
event, possibly a merger. 

\begin{figure}  
\begin{center}
{\leavevmode
\epsfxsize=.48\textwidth \epsfbox{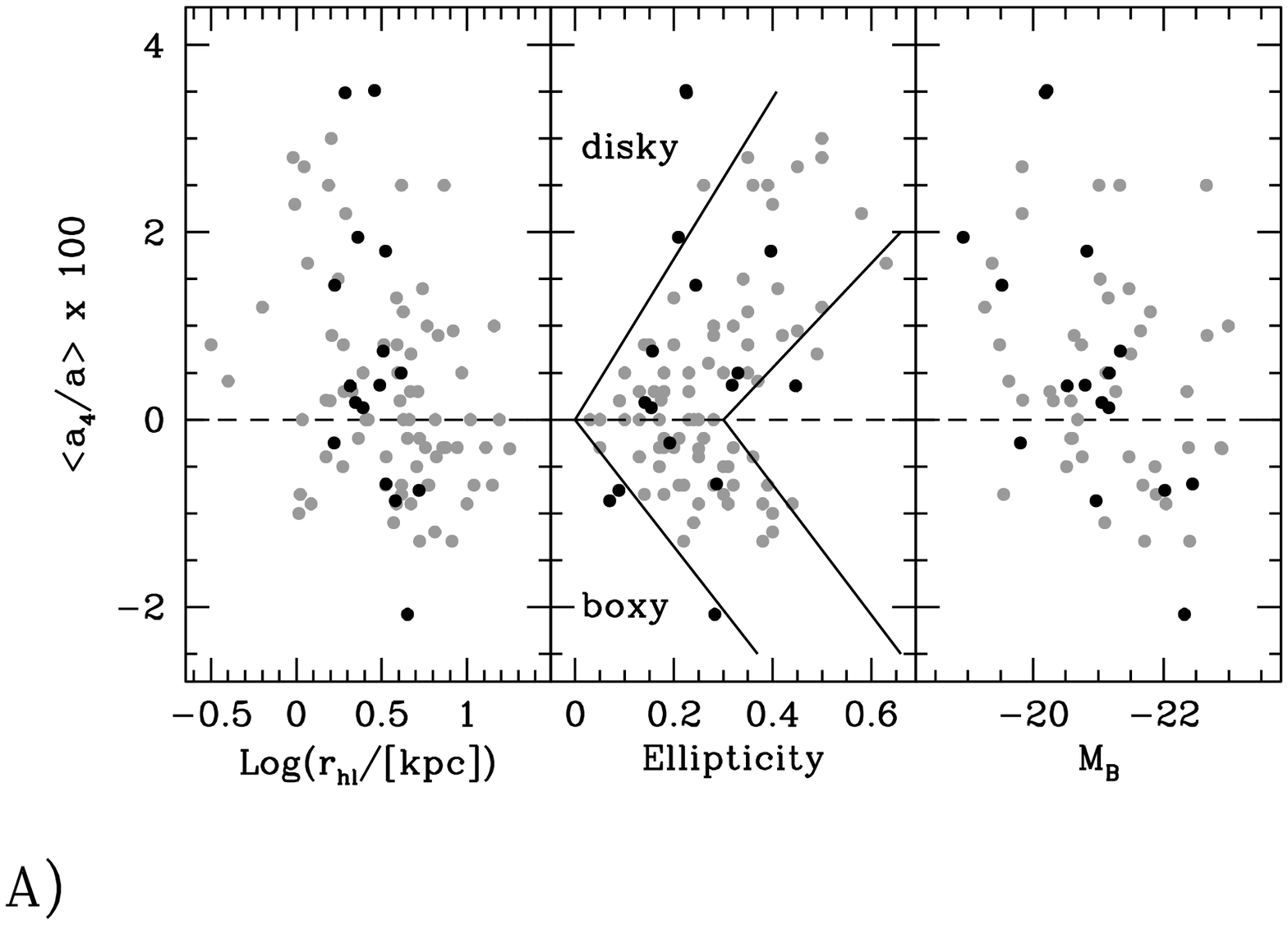} \hfil
\epsfxsize=.48\textwidth \epsfbox{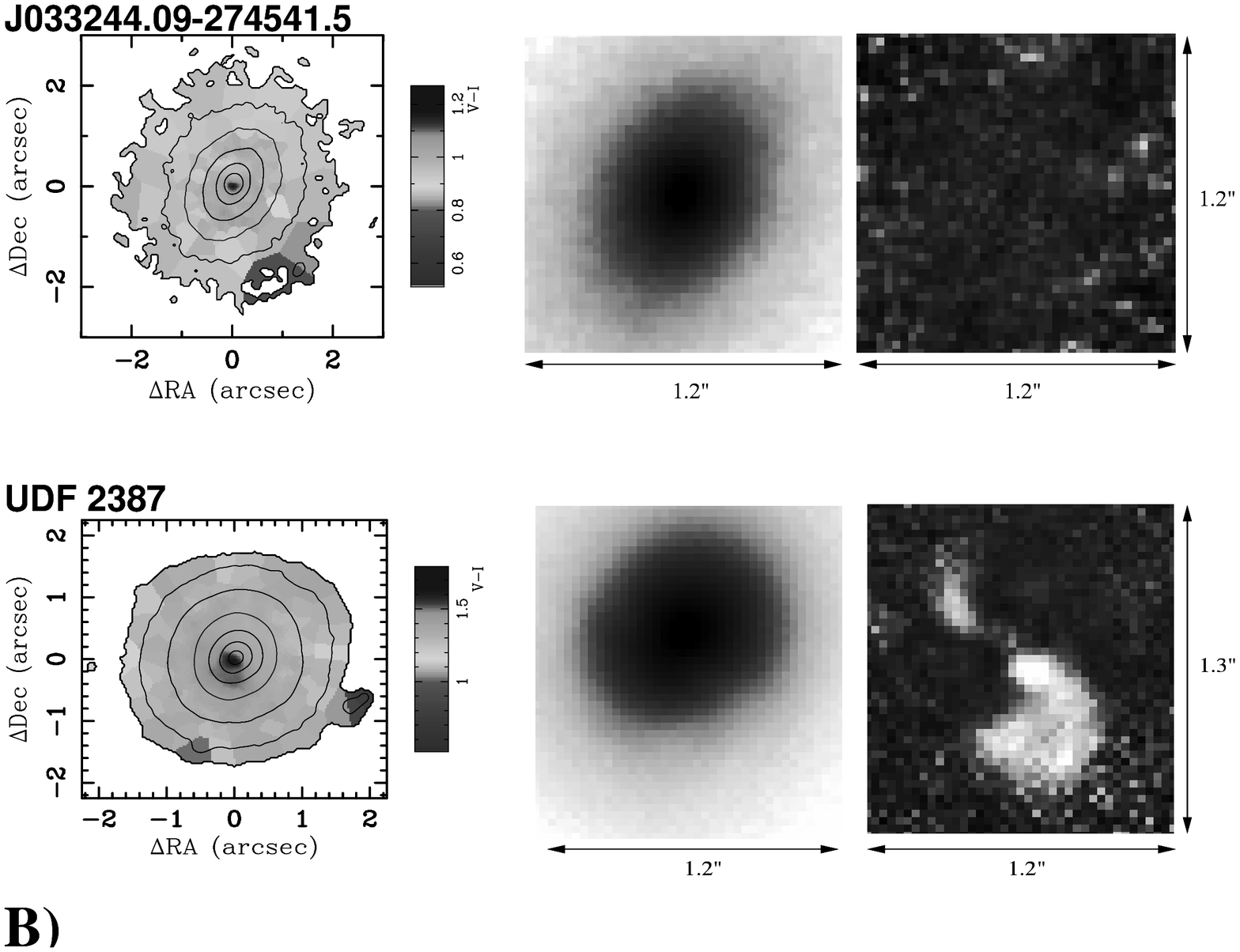}
}
\end{center}
\caption{{\bf Detailed structure from the UDF.} {\sl LEFT:} The
structural parameter $\langle a_4/a\rangle$ measures the diskyness/boxyness of a
galaxy. The black dots are UDF early-type galaxies, which are very
similar to the local sample of Bender et al. (1989; grey dots) {\sl
RIGHT:} Two early-type galaxies from the UDF images. The left panels show
the $V-i$ colour maps; the middle panels give the best fit to a smooth
surface brightness profile (in the $i$ band). The right panels give
the residuals. Both galaxies have red cores and do not show any
appreciable photometric departure from a ``standard'' early-type
galaxy.}
\label{fig:results}
\end{figure} 

\section{Assembling the puzzle} 
We have assembled a sample of intermediate-redshift galaxies which --
by construction -- already look like today's elliptical galaxies. Such
approach has the advantage of not relying on colour selection which
will bias the sample in favour of the old stellar populations seen in
local early-type systems. Given the depth and spatial resolution of
the ACS images, this sample robustly comprises all galaxies which can
be expected to evolve into local early-type systems.  One could still
argue that the ``filter'' applied regarding the Kormendy relation will
bias against other possible progenitors. Unfortunately, such an analysis
cannot rely on a morphological estimation. Furthermore, the small
scatter of the colour magnitude relation at $z=0$ (e.g. Terlevich,
Caldwell \& Bower 2001) implies that blue galaxies in the redshift
range pertinent to this sample, $z\simlt 1$, are unlikely progenitors
of today's elliptical galaxies. The evidence gathered from this sample
presents morphologically-selected early-type galaxies as mostly made
up of an old, passively evolving population, where the colour
gradients are purely caused by a range of metallicities. The fraction
of blue galaxies increases with redshift and corresponds to fainter
systems (i.e. 'downsizing'), but stays overall at a low fraction,
around 20\%. On the other hand, the comoving number density presents a
significant decrease with redshift, suggesting either a strong
evolution of the assembly history as predicted by hierarchical models,
or a selection bias caused by dust.

\vfill 
\end{document}